\documentclass[aps,prb,twocolumn,superscriptaddress]{revtex4}
\usepackage{amssymb}
\usepackage{amsmath}
 
\usepackage[pdftex]{graphicx}

\usepackage{amsfonts}
\usepackage{bbold}
\usepackage{color}

\begin{document}
\title{Spin-polarized tunneling into helical edge states: asymmetry and conductances}
\author{D.N. Aristov}
\affiliation{NRC ``Kurchatov Institute", Petersburg Nuclear 
Physics Institute, Gatchina 188300, Russia}
\affiliation{St.Petersburg State University, 7/9 Universitetskaya nab., St. Petersburg, 199034 Russia}
\author{R.A. Niyazov}
\affiliation{NRC ``Kurchatov Institute", Petersburg Nuclear 
Physics Institute, Gatchina 188300, Russia}
 
\begin{abstract}
We consider tunneling from the spin-polarized tip into the Luttinger liquid edge state of quantum spin Hall system. This problem arose in context of the spin and charge fractionalization of an injected electron.  Renormalization of the dc conductances of the system is calculated in the fermionic approach and scattering states formalism. In lowest order of tunneling amplitude we confirm previous results for the scaling dependence of conductances. Going beyond the lowest order we show that interaction affects not only the total tunneling rate, but also the asymmetry of the injected current.  The helical edge state forbids the backscattering, which leads to the possibility of two stable fixed points in renormalization group sense, in contrast to the Y-junction between the usual quantum wires.
\end{abstract}

\maketitle

\section{Introduction}
    
Topological insulators (TI) is a class of materials  apparently important for future electronic devices.~\cite{Hasan2010,Qi2011} Being an insulator in the bulk, the TI necessarily has conducting states on its edge. The spin-orbit interaction leads to electrons spin-momentum locking in the edge states creating so called helical states.  Backscattering on the nonmagnetic impurities in the same channel of such edge states is absent due to TI time-reversal symmetry. It allows a creation of devices with a lossless electron transport.

Electrons on the edge of 2D TI are confined to one spatial dimension (1D). One dimensional electron transport near the Fermi energy is described by well-known Tomonoga-Luttinger liquid (TLL) model.  \cite{GiamarchiBook} The TLL  corresponds to strongly correlated electron matter, and the 1D helical edge states add new peculiarities to it.
This explains large experimental and theoretical interest to helical TLLs, with several suggested ways of probing 
their strongly correlated nature.
Investigating the helical edge states  was theoretically proposed by means of magnetic field   in~\cite{Braunecker2012,Kharitonov2012,Soori2012}, by specific spatial construction of helical state in~\cite{Dolcetto2013a,Calzona2015a,Calzona2016a} and by quantum dots  in~\cite{Chao2013,Dolcetto2013}.

Another natural way of  probing -- tunneling from the tip into the edge state -- attracted a lot of attention.
There is experimental evidence~\cite{Steinberg2007} and theoretical explanation~\cite{Hur2008} 
that   the electron injection even into usual TLL may be characterized by charge fractionalization resulting in the currents imbalance.  For helical TLL probing by the tip  was considered theoretically in papers~\cite{Pugnetti2009,Das2011,Garate2012,Khanna2013,Calzona2015,Santos2016}.
Experimentally it was confirmed  that  TI edge state is a quantum spin Hall state~\cite{Kim2014} , helical behavior of the edge state was observed in~\cite{Li2015}, the STM experiments were reported in~\cite{Pauly2015,Wu2016}.

The transport properties of TLL model may be studied in two main approaches. The bosonization approach~\cite{GiamarchiBook}  takes into account the electron interaction exactly, the impurities are regarded as perturbation. The fermionic approach treats interaction perturbatively  and studies the renormalization of  impurities/junctions  in $S$-matrix formalism of asymptotic electronic states.  
Transport in quantum point contact is described by the conductances (transparencies) of junctions. The problem with one impurity in   quantum wire can also be interpreted as the point contact of two wires. The latter case    was considered within the TLL model in bosonization~\cite{Giamarchi1988,Furusaki1993,Kane1992} and fermionic~\cite{Matveev1993a,Yue1994} approaches. The conclusion of both approaches is that repulsive interaction renormalizes bare conductance to zero (total reflection case) whereas  the attractive interaction leads to perfect transmission   in the low temperature limit. 

Further studies of usual TLL in bosonization approach addressed Y-junction with equal interaction constant~\cite{Oshikawa2006} and arbitrary interaction in three wires~\cite{Hou2012}. The corner junction of helical TLLs was considered for  two TI's  in ~\cite{Hou2009,Teo2009,Stroem2009}. 

In the fermionic approach the renormalization group (RG) equations for $S$-matrix characterizing the junction for arbitrary  number of quantum wires  was presented  in ~\cite{Lal2002,Das2004}.  Conductances of Y and X junctions were  analyzed in the first order of interaction.
 
A partial summation of the perturbation series was performed in \cite{Aristov2009}  in order to recover the exact  scaling exponents for conductance for one impurity. The results were 
used for the RG analysis of  Y-junction ~\cite{Aristov2011,Aristov2012a,Aristov2013}
and 
generalized to arbitrary  number of wires ~\cite{Aristov2012}. 
The contact of two helical TLL and tunneling from unpolarized spinful wire  into  the  helical edge state were considered  in ~\cite{Aristov2016a}. The agreement between bosonization and fermionic approaches has been obtained when the comparison was available.   
 
\begin{figure} 
\includegraphics[width=0.5\columnwidth]{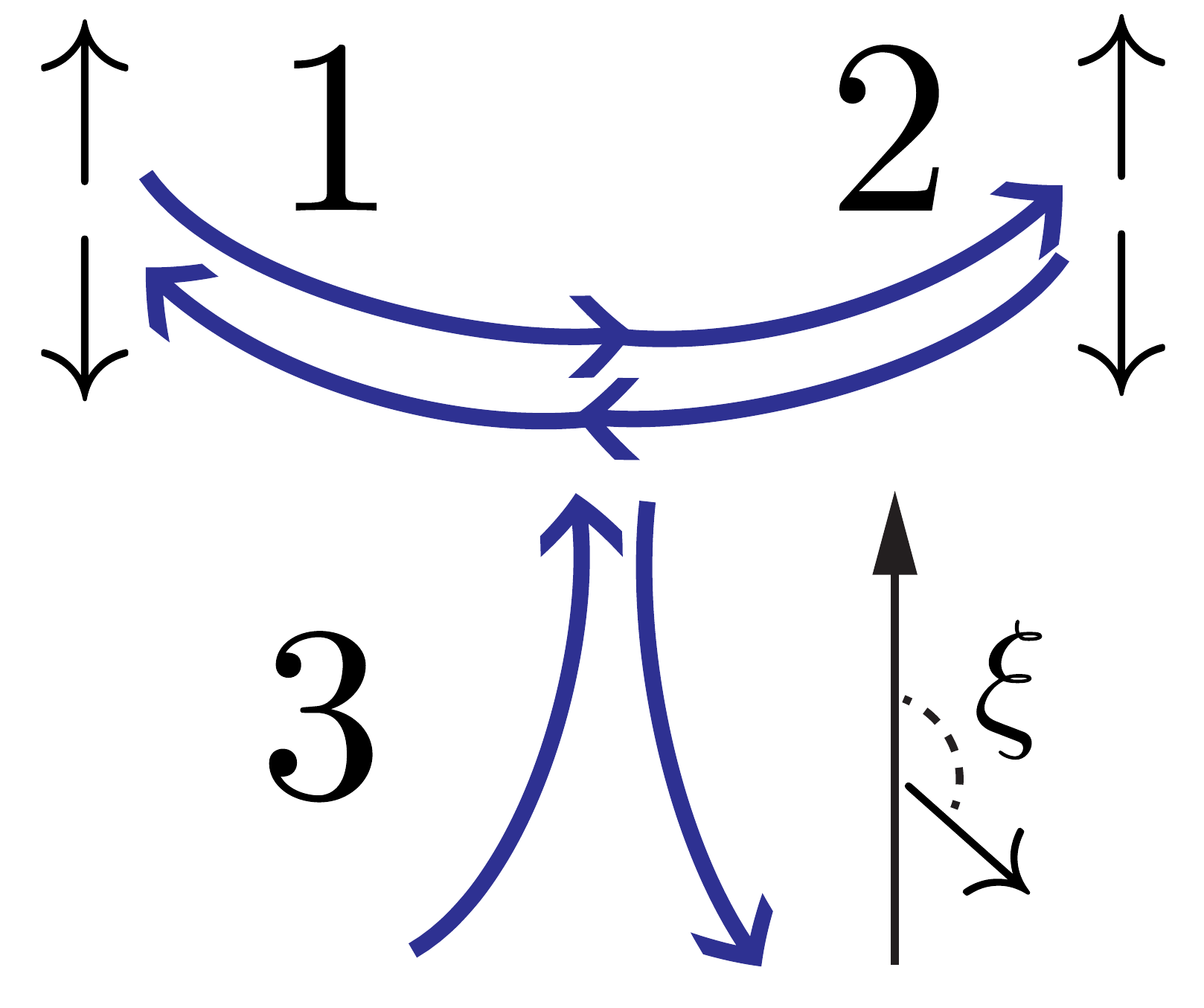} 
\caption{\label{fig:setup}
Schematically shown scanning tunneling microscopy of the helical edge state of topological insulator by the spin-polarized tip. $\xi$ is an angle between electron spin in the tip and spin quantization axis of the helical state.}
\end{figure}

In the present letter we focus on the spin-polarized tunneling into the helical edge state. This problem was previously studied within bosonization approach~\cite{Das2011}, with the setup consisting of a point contact between fully spin-polarized tip and the helical edge state (see Fig.\ref{fig:setup}). Generally, spin polarization in the tip differs from the spin direction of the edge state fermions by some angle, $\xi$. In the absence of interaction this angle defines the left-right current asymmetry of the injected current. The charge fractionalization was demonstrated for the injection of electrons even with polarization parallel to the edge states in case of infinite helical TLL without Fermi liquid leads.  Generalization of this result to time-dependent injected pulses was done in~\cite{Garate2012}. The effect of metallic leads for the visibility of fractionalization phenomena in transport measurement in helical TLL was analyzed in ~\cite{Calzona2015}. It was argued there that the fractionalization cannot be observed for dc currents, while information about helical TLL physics can be extracted from studying the transient dynamics. 

In our approach below  we use the fermionic formalism of scattering states, which implies the existence of Fermi liquid leads. It results in the absence of charge fractionalization in dc limit for the parallel polarization of the tip.  For non-parallel polarization we show that the left-right asymmetry of the dc current, injected into helical TLL, is not defined entirely by the polarization angle but is  renormalized along with the total tunneling conductance,  with the resulting coupled RG  equations.  The asymmetry renormalization appears in first order of interaction strength and first order of tunneling conductance, and we discuss it in detail.  Simultaneous renormalization of two quantities is tied to the existence of saddle point RG fixed point (FP) in the phase portrait of the system. \cite{Aristov2010}  In the absence of backscattering in the helical TLL,  this non-universal FP separates two truly stable FPs.  This in turn may lead to qualitatively different renormalization of asymmetry and the total tunneling rate for different interaction strength. 

\section{Theoretical formalism
\label{sec:model}}

\subsection{The model
\label{subsec:model}}

We consider Y-junction of the quantum wires shown in Fig.~\ref{fig:setup}. Two quantum wires  labeled by $j=1,2$ correspond to the TI helical edge state   and the wire with $j=3$ is the tunneling tip with fully polarized electrons. The polarization vector of electrons in the tip is different from the quantization axis in the TI edge state.  We denote this difference by the angle $\xi$, so that $\xi=0(\pi)$  describes the tip polarization parallel (antiparallel) to the right-moving fermions in the edge state.   

The helical state of the main wire   is described by 
the TLL model with the interaction between electrons of short-range forward-scattering type. 
All three quantum wires of length $L$ are adiabatically connected to the reservoirs or leads such that additional scattering is absent. 
Scattering electron states flowing in and out the junction are connected by the $S$ matrix.

TLL model Hamiltonian with linearized spectrum near the Fermi energy and in the scattering states formalism reads
\begin{equation}
\begin{aligned}
\mathcal{H} &=\int_0^{\infty}dx[H^{0}+H^{int} \Theta( \ell < x < L )]\,,   \\
H^{0} &= v_{F} \Psi_{in}^{\dagger }i\nabla \Psi _{in}-v_{F}\Psi
_{out}^{\dagger }i\nabla \Psi _{out}\,, \\
H^{int} &= 2\pi v_{F}  \sum\limits_{j=1}^3  g_{j}
 \widehat{\rho }_{j} \widehat{\widetilde{\rho }}_{j} \,.  
\end{aligned} 
\label{Ham}
\end{equation}
Density operators are defined by $\widehat{\rho }_{j,in}=\Psi ^{+}\rho _{j}\Psi = \widehat{\rho }_{j}$, 
and $\widehat{\rho }_{j,out}=\Psi ^{+}\widetilde{\rho }_{j}\Psi =\widehat{\widetilde{\rho }}_{j}$, 
where $\widetilde{\rho }_{j}=S^{+}\cdot \rho _{j}\cdot S$ 
and the density matrices are given by $(\rho _{j})_{\alpha \beta }=\delta _{\alpha \beta }\delta _{\alpha j}$ and 
$(\widetilde{\rho }_{j})_{\alpha \beta }=S_{\alpha j}^{+}S_{j\beta }$. Electron interaction constant in the edge state is defined by $g_1=g_2=g$ and $g_3=0$ in the tip ; we set $v_F=1$ below. Here we introduce window function $\Theta( \ell < x < L )$ such that $\Theta(x)=1$, if $\ell < x < L$ and zero otherwise.  The ultraviolet cutoff  at $x = \ell$  is required for our model point-like interaction, $H^{int}$,  and for finite-range interaction $\ell$ is associated with the screening length, see Appendix~\ref{app:cutoff}.

Ingoing and outgoing electronic waves $\psi_{j,\text{in}}, \psi_{j,\text{out}}$ are connected at the junction  by the $S$ matrix $\Psi _{out}(x)=S\cdot \Psi
_{in}(x)$ at $x\to 0$, where  $\Psi _{in(out)}=(\psi _{1,in(out)},\psi _{2,in(out)},\psi _{3,in(out)})$.
The $S$-matrix characterizes the scattering in the junction and belongs to the unitary group $U(3)$. 
Our setup implies the chiral property, $|S_{13}| = |S_{32}|$,  $|S_{23}|= |S_{31}|$.
Due to the  time-reversal invariance of TI,  the backscattering in the edge state  without the tip is forbidden.  The same is true in case in which the polarization of the tip is parallel to the spin direction of right or left mover, $\xi=0, \pi$.   

In general, the chiral S-matrix is characterized by three parameters \cite{Aristov2013}.
The absence of backscattering reduces the appropriate $S$ matrix to the following two-parametric form   
\begin{equation}\label{eq:smatrix}
 \begin{aligned}
  S&=  \begin{pmatrix}
r_1& t_{12} & t_{13} \\
t_{21} &  r_1 & t_{23} \\
 t_{31} &  t_{32} &r_2 \\
\end{pmatrix} \\
 r_1&=\sin ^2\frac{\theta }{2}\sin \xi  , \quad r_2= \cos \theta , \\
 t_{12}&=-\cos \theta \cos ^2 \frac{\xi }{2}-\sin ^2\frac{\xi }{2}, \quad t_{21}=t_{12} |_{\xi \rightarrow \pi -\xi }, \\
 t_{13}&=t_{32}=\cos \frac{\xi }{2} \sin \theta , \quad t_{23}=t_{31}= \sin \frac{\xi }{2} \sin \theta . 
\end{aligned}
\end{equation}
One parameter, $\theta$, is responsible for the tunneling  amplitude and $\theta=0$ corresponds to the fully detached tip, in this case  there is a perfect transmission through the edge state.  Second parameter $\xi$ is the above angle between electron spin in the tip and spin quantization axis of the helical state (see Appendix~\ref{app:rot}).

\subsection{Reduced conductances}  
  
In the linear response regime the conductances are defined by $I_{i}=C_{ik} V_k$ where $I_j$ and $V_j$ are 
the current flowing towards in the junction and the electric potential in $j$th lead, respectively. 
In the dc limit one easily obtains from Kubo formula $C_{ij}=\delta_{ij}-Y_{ij} $, with $Y_{ij} = |S_{ij}|^{2}$.  The total current conservation and the absence of current for equal voltages lead to conditions $\sum _{i}C_{ij} = \sum _{j}C_{ij} = 0$, which also stem from the unitarity of $S$-matrix.  This means that we can introduce new current  and voltage combinations to show the condition explicitly. 
\cite{Aristov2011a}

We choose new combinations according to relations $I_{i}^{\text{new}} = \sum_{k} R_{ki} I_{k}$,  $V_{i}^\text{new} = \sum_{k} R_{ki} V_{k}$, where the matrix
\begin{equation}\label{eq:r}
\mathbf{R}=
\begin{pmatrix}
 \frac{1}{\sqrt{2}} & \frac{1}{\sqrt{6}} & \frac{1}{\sqrt{3}} \\
 -\frac{1}{\sqrt{2}} & \frac{1}{\sqrt{6}} & \frac{1}{\sqrt{3}} \\
 0 & -\sqrt{\frac{2}{3}} & \frac{1}{\sqrt{3}} 
\end{pmatrix}
\end{equation}
has   properties $\mathbf{R}^{-1}=\mathbf{R}^{T}$, $\mbox{det }\mathbf{R}=1$.

The corresponding matrix of conductances $\mathbf{C}^R=\mathbf{R}^T\,\mathbf{C}\, \mathbf{R}$ for the above form of $S$, Eq. \eqref{eq:smatrix},  becomes 
\begin{equation}
\mathbf{C}^R =
\begin{pmatrix}
1-a &- c& 0 \\
 c &1- b& 0\\
 0 & 0 & 0 
\end{pmatrix}.
\end{equation}
with 
\begin{equation}
\begin{aligned}\label{eq:abc}
a&=\tfrac 12 (1+\cos^2 \theta ) \cos^2 \xi- \cos \theta \sin^2 \xi, \\
b&=\tfrac 14 (1+3 \cos 2\theta), \\
c&=\tfrac{\sqrt{3}}{2} \cos \xi \sin^2 \theta.
\end{aligned}
\end{equation}
It is convenient to analyze  the quantity $\mathbf{Y}^R =1 - \mathbf{C}^R$ below. Clearly, only two components of $\mathbf{Y}^R$ are independent.   The chiral component of conductance, $c$, depends on the asymmetry parameter, $\cos \xi$, which includes the polarization angle, $\xi$.  

We can also introduce more ``physical'' combinations of currents, $\mathbf{I}^\text{phys} = (I_a, I_b)$, and voltages, $\mathbf{V}^\text{phys}= (V_a, V_b)$,  as
\begin{equation}
\begin{aligned}\label{eq:physcurrents}
I_a &= \tfrac 12 (I_1-I_2)\,,  \quad &I_b= \tfrac 13(I_1 + I_2 -2 I_3)  \,, \\
V_a &= V_1-V_2 \,,    \quad &V_b = \tfrac 12(V_1 + V_2 -2 V_3) \,.
\end{aligned}
\end{equation}
In terms of these, the matrix of conductances $\mathbf{I}^\text{phys}=\mathbf{G} \mathbf{V}^\text{phys}$ is given by 
\begin{equation}\label{eq:physcond}
\mathbf{G}=
\begin{pmatrix}
G_a & G_{ab}\\
 G_{ba} & G_b
\end{pmatrix}=
\begin{pmatrix}
\frac 12 (1-a) &- \frac{c}{\sqrt{3}}\\
\frac{c}{\sqrt{3}}& \frac 23 (1-b)
\end{pmatrix}.
\end{equation}
with the property, $0\le G_{a,b} \le1$, see \cite{Aristov2011a}.

\subsection{Renormalization group equations}

The dc transport through the chiral Y-junction was studied in fermionic formalism in \cite{Aristov2012a,Aristov2013}. We sketch here the derivation of the corresponding formulas.   
The principal first order  interaction correction to the conductances reads~\cite{Aristov2012a}
\begin{equation}
C_{jk}=C_{jk}\big\rvert_{\mathbf{g}=0}
+\tfrac{1}{2}\sum\limits_{l,m} \mbox{Tr}\left[\widehat{W}_{jk}\widehat{%
W}_{lm} \right] g_{ml}\Lambda \,,
\label{Rgorder}
\end{equation} 
where we defined  $C_{jk}\rvert_{\mathbf{g}=0}=\delta _{jk}-Y_{jk}$,  a matrix $g_{ml}=g_l \delta_{ml}$ with  interaction constants $g_l$   in the Hamiltonian~\eqref{Ham}, nine  $3\times 3$ matrices  $\widehat{W}_{jk}= [\rho_{j}, \widetilde\rho_{k}]$ and the 
trace operation $\mbox{Tr}$ is defined with respect to the 3$\times$3 matrix space
of $\widehat{W}$'s. 
The scale-dependent term $\Lambda = \ln (\min[L, \xi ]/\ell )$  with  the temperature correlation length $\xi = v_{F}/T$.
 
The above correction is the leading contribution, $\sim g \Lambda$, to conductance. It was shown \cite{Aristov2011a} that the higher-order corrections obey the scaling hypothesis for the conductance and are generated by a set of the differential RG equations. In its simplest form,  these equations are obtained 
   by differentiating Eq.  \eqref{Rgorder} over $\Lambda$ (and then putting $\Lambda =0$) which gives
\begin{equation}
\frac{d}{d\Lambda }Y_{jk}^{R}=-\frac{1}{2}\sum \limits_{l,m}
\mbox{Tr}\left[\widehat{W}_{jk}^{R}\widehat{W}_{lm}^{R}\right] g_{ml}^{R} \, ,
\label{eq:RGgeneral}
\end{equation}
where all quantities with superscript $R$ are defined as $\mathbf{A}^R=\mathbf{R}^T\,\mathbf{A}\, \mathbf{R}$.

The main source of  linear-in-$\Lambda$ subleading corrections corresponds to a ladder series of diagrams, which can be summed analytically  \cite{Aristov2011a}. These corrections are independent of  the scheme of regularization in RG procedure and lead to the scaling exponents for conductances, coinciding with those obtained by bosonization method.  

The result of the ladder summation~\cite{Aristov2012,Aristov2013} is the RG equation similar to above Eq.\  \eqref{eq:RGgeneral} with the replacement  
\begin{equation}\label{eq:gladder}
\mathbf{g} \to \bar{\mathbf{g}}=2(\mathbf{Q}-\mathbf{Y})^{-1}\,.
\end{equation}
The matrix $\mathbf{Q}$ depends on the Luttinger parameters $K_j$ of individual wires and has the form
\begin{equation}
\begin{aligned}\label{eq:qdef}
Q_{jk}& = q_{j}\delta _{jk}\,,\quad q_{j} =(1+K_{j})/(1-K_{j})\, , \\
K_{j}&=[(1-g_{j})/(1+g_{j})]^{1/2}.
\end{aligned}
\end{equation}
In our case we have $g_1=g_2=g$ and $g_3=0$. The latter equality corresponds to $q_3 \to \infty$, whose limit   is easily taken in \eqref{eq:gladder}.

Now we are in a position to obtain RG equations of the conductances. However, the equivalent  RG equations in terms of parametrization~\eqref{eq:smatrix} have  simpler form, and we choose it below. After some algebra we obtain from \eqref{eq:abc}, \eqref{eq:RGgeneral},  \eqref{eq:gladder} the following set of  equations   
\begin{equation} 
\begin{aligned}
\frac{d \theta}{d \Lambda}  &= -\frac18 (1-K)\sin \theta  \frac{  F_1(\theta)+ F_2 (\theta) \cos 2\xi}{  F_3(\theta) D(\theta, \xi)} , \\
\frac{d \xi}{d \Lambda}  &=\frac14 (1-K)\frac{(1-\cos \theta) \sin 2\xi}{  D(\theta,\xi)},\\
F_1&=2(1-K) - (3-K) \sin^2 \frac \theta2 
\\  & - 3 (1-K) \sin^2 \theta, \\
F_2 &= \sin^2 \frac \theta 2 (2 - 3 (1-K)  \sin^2 \frac \theta 2 ), \\
F_3 &= 1-(1-K)  \sin^2 \frac \theta 2, \\
D(\theta,\xi) &= K + (1-K) \sin^{2}\frac\theta2 \left( 1-\sin^{2}\frac\theta2 \sin^{2} \xi  \right).
\end{aligned} 
\label{eq:rgtx}
\end{equation}
These equations are analyzed in the next section.  Notice that the asymmetry parameter $p=\cos \xi$ is now determined not only by the initial polarization angle, but also by the interaction strength and tunneling amplitude.

\section{Spin-polarized tunneling \label{sec:sptip}}

\subsection{Lowest order calculation}

Let us first consider the limit of weak tunneling into the helical edge state. This means that both tunneling from the tip and the backscattering in the edge state due to this tunneling are approximately zero. In terms of the $S$-matrix \eqref{eq:smatrix} we write $|r_2|^2\simeq  1$ and $|r_1|^2\simeq 0$. First condition gives $\theta\simeq 0$ or $\theta\simeq \pi$, and the second condition states that if $\theta \simeq\pi$ then $\xi \simeq 0$ or  $\xi \simeq\pi$. These three seemingly unconnected regions form a neighborhood of the fixed point (FP) A, as is discussed below. 

Confining ourselves to the  linear-in- $\theta$ terms, we get   from Eqs.~\eqref{eq:rgtx}:
\begin{equation}\label{eq:rgtheta1}
\begin{aligned}
\frac{d \theta}{d \Lambda} & = - \frac{(1-K)^2}{4K} \theta\, , \qquad
\frac{d \xi}{d \Lambda}  = 0 \,.
\end{aligned}
\end{equation}
Hence, the parameter $\theta$, which is related to the barrier transparency, is renormalized to zero with the scaling law $\theta = \theta_0 \exp(-\nu \Lambda / 2)$, and the tunneling exponent $\nu = (1-K)^2/2K$. The asymmetry ratio $ t_\downarrow / t_\uparrow = \tan \xi/2$ is not renormalized.

This result is in exact agreement with the earlier work \cite{Das2011}, where the renormalization of the tunnel amplitudes $t_\uparrow$ and $t_\downarrow$  was studied in the bosonization approach.  
Equation (9) of the above paper obtains the currents flowing to the right $I_{\text{R}}$ and to the  left $I_{\text{L}}$ ends of the edge state. One has   
\begin{equation}\label{eq:currents_Das}
\begin{aligned}
I_{\text{L}}+I_{\text{R}}&=I_0\,, \qquad 
I_{\text{L}}-I_{\text{R}} =-K I_0 \cos \xi \,.
\end{aligned}
\end{equation}
In our notation ~\eqref{eq:physcurrents}
the current $I_{0}$ equals to $I_{b}$, and the difference of the currents $I_{\text{L}}-I_{\text{R}}$ is $2I_{a}$.  For zero bias in the edge state, $V_a=0$, we have    $I_b= G_b V_b$,  $I_a=G_{ab} V_b$ and   obtain 
  from Eqs.\  \eqref{eq:abc}, \eqref{eq:physcond} : 
\begin{equation}
2 I_a=-I_b \cos \xi
\end{equation}
This equation coincides with  \eqref{eq:currents_Das}   apart from the factor $K$, which is absent in dc limit in our setup with the Fermi leads. The absence of this factor in the similar setup  was also obtained  in the work ~\cite{Calzona2015} in bosonization approach.

Now we consider the next terms in the expansion of RG equations in  $\theta$. Quadratic-in $\theta$ term appears only in RG equation for asymmetry angle:
\begin{equation} \label{RGxi}
\frac{d \xi}{d \Lambda}  = \frac{1-K}{8K} \theta^2 \sin 2 \xi.
\end{equation}
Solution of the equation shows that the asymmetry angle is renormalized according to the law 
\begin{equation}\label{eq:polarization1}
\tan \xi = \tan \xi_0 e^{-(\theta^2-\theta_0^2)/2(1-K)}.
\end{equation}
Since $\theta$ is renormalized to zero (see eq.~\eqref{eq:rgtheta1}), we find the whole renormalization of   $\xi$ insignificant for small bare tunneling amplitude  $\theta_0$. At the same time the renormalization of $\xi$ in \eqref{RGxi} is  first order interaction effect for small interaction $\frac{1-K}{K}\approx g \ll 1$,  whereas the tunneling exponent for $\theta$ in Eq.\ \eqref{eq:rgtheta1} is given by the second order of interaction, $\frac{(1-K)^2}{K}\approx g^2$.

\begin{figure} 
\includegraphics[width=0.98\columnwidth]{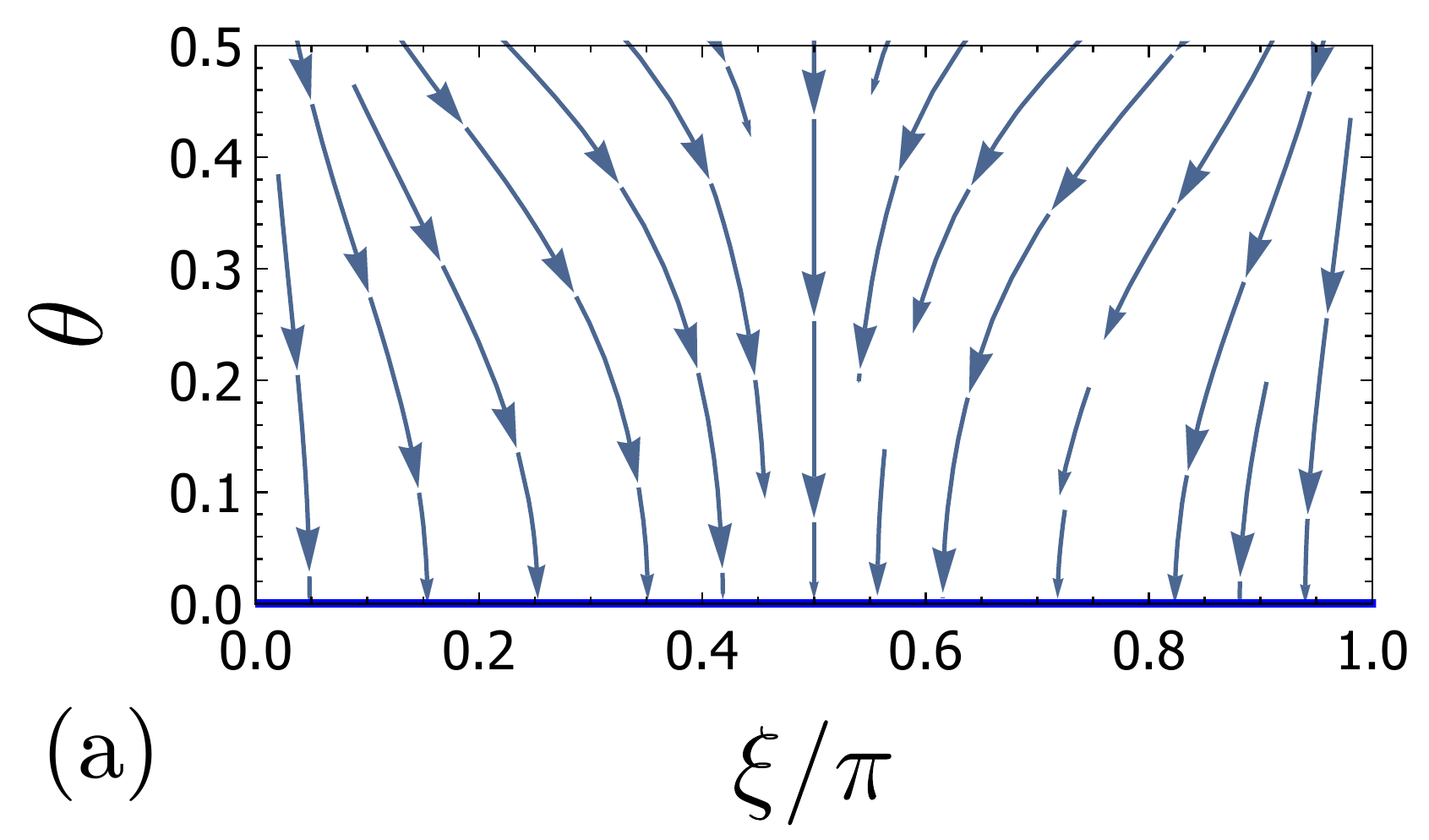} 
\includegraphics[width=0.94\columnwidth]{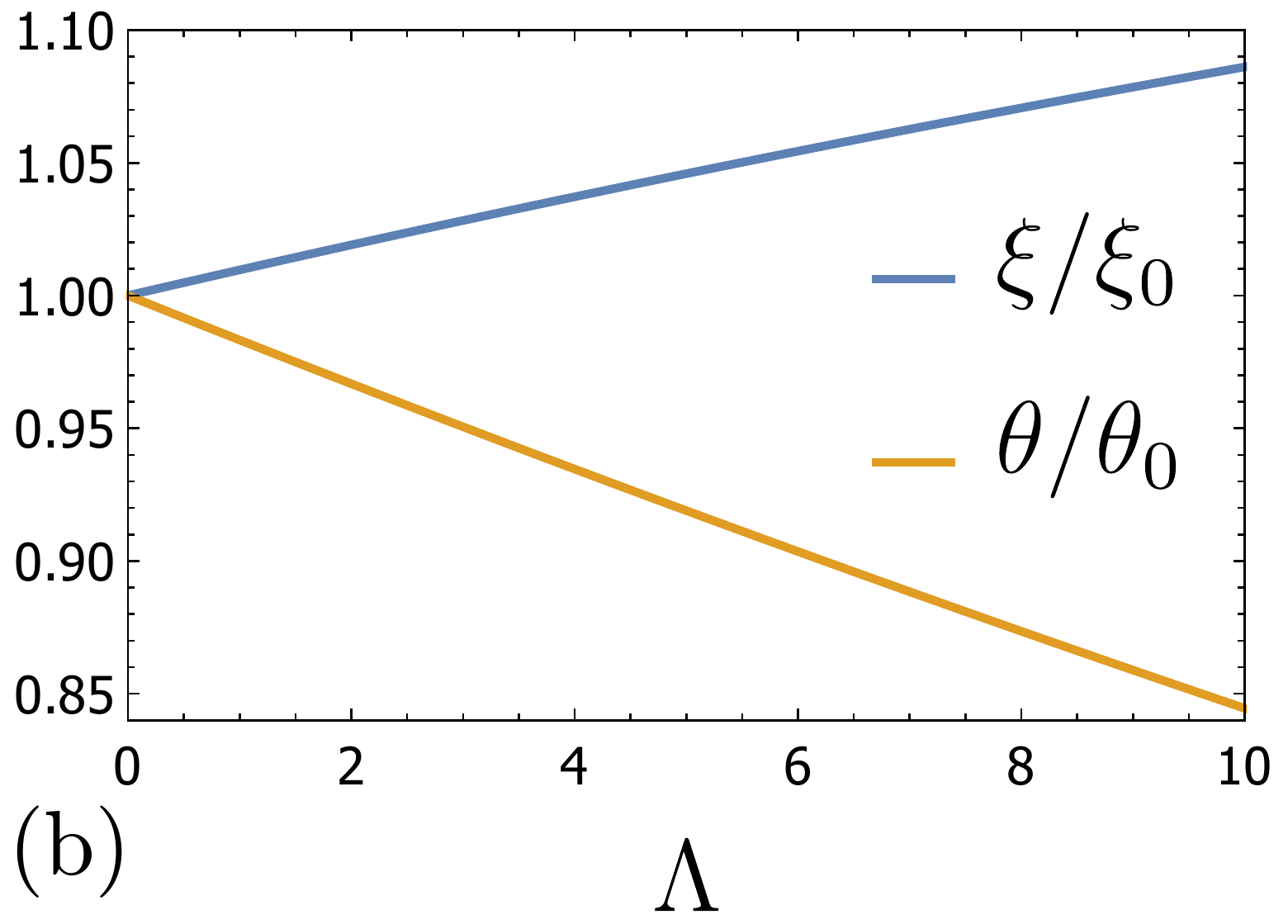} 
\caption{\label{fig:approx2}
The panel (a) shows RG flows in the plane $(\xi, \ \theta)$ for $g=0.26$ ($K=0.77$).  
Relative renormalization of $\xi$ and $\theta$  for the
flow with $\xi_{0}=0.3$, $\theta_{0}=0.4$  is shown in panel (b). }
\end{figure}

We see that if we discard the small-in-$\theta$ terms in RG equation for $\xi$, then we do not obtain renormalization of asymmetry at all. However the smallness-in-$\theta$ may be compensated by the less power of interaction in the RG flow of $\xi$. As a result for small interaction and tunneling amplitude the renormalization of $\theta$ and $\xi$ may be of the same order as demonstrated in Fig.~\ref{fig:approx2}.

We also note here that Eq. ~\eqref{eq:polarization1} can be rewritten in terms of the asymmetry $p = \cos \xi=-2 I_a / I_b$ as 
\begin{equation}
\frac{d p}{d \Lambda}  =- \frac{1-K}{4K} \theta^2 p (1-p^2).
\label{eq:polarization2}
\end{equation}
This equation possesses three fixed points (FPs) $p=0,\pm 1$, and if $\theta^2$ were non-vanishing at $\Lambda \to \infty$, then we would obtain only $p=0$ as a stable FP for repulsive   interaction, $K<1$. In the leading order of interaction, we can combine first equation in \eqref{eq:rgtheta1} and \eqref{eq:polarization2} to represent the RG trajectory in the form 
\begin{equation}\label{eq:rgpolarization1}
\theta^2 - g \log \frac{p^2}{1-p^2} = \mbox{const}. 
\end{equation}

\subsection{M point}

The terms of the order $\theta^3$ appear only in the equation for  $\theta$    :
\begin{equation}\label{eq:rgtheta3}
\frac{d \theta}{d \Lambda} = - \frac{(1-K)^2}{4K} \theta + \frac{ (1-K) (\kappa -\cos 2\xi)}{16 K} \theta^3,
\end{equation}
where $ \kappa \equiv (1-K) (2/3+K)+1/K$. For small interaction $\kappa \approx 1$ and the equation \eqref{eq:rgtheta3} becomes 
\begin{equation}
\frac{d \theta}{d \Lambda} = - \frac{g^2}{4}\theta+ \frac{g (1-\cos 2\xi)}{16} \theta^3.
\end{equation}
First, one can see here that   the $\theta^3$ -term is unimportant  in parallel polarization case, $\xi=0,\pi$; the right-hand side of Eq.\ \eqref{RGxi} is zero. Thus the lines $\xi=0,\pi$ are RG fixed lines. 

Otherwise the  combination of \eqref{eq:rgtheta3} with \eqref{RGxi} reveals the existence of  
  non-universal fixed point M at $\xi_M=\pi/2$,  $\theta_M \approx \sqrt{2g}$.  Similar to above Eq.\ \eqref{RGxi}, the next-order term in $\theta$ expansion contains smaller power of interaction strength. This situation was discussed   for non-chiral symmetric Y-junction in \cite{Aristov2010}. 

Solving  RG equations for asymmetry $p$ and $\theta$ gives for the RG trajectory
\begin{equation} \label{eq:rgpolarization2}
\theta^2 p - g \log \frac{1+p}{1-p} = \mbox{const} \,.
\end{equation}
Despite the apparently different form of Eqs.\  \eqref{eq:rgpolarization2} and \eqref{eq:rgpolarization1}, these equations yield numerically close curves for small $\theta < \theta_{M}$. Indeed the existence of the next term in equation~\eqref{eq:rgtheta3} does not seriously affect RG flows in this limit.

Already at the level of approximate Eqs.\  \eqref{eq:rgtheta3}, \eqref{RGxi} one can expect that  for $\theta \approx \theta_M$ the RG flow for the asymmetry $\cos \xi$ may exceed  the RG flow of barrier transparency $\theta$.  We discuss this point in more detail in the next subsection.

\subsection{Non-perturbative RG results}

\begin{figure} 
\includegraphics[width=0.99\columnwidth]{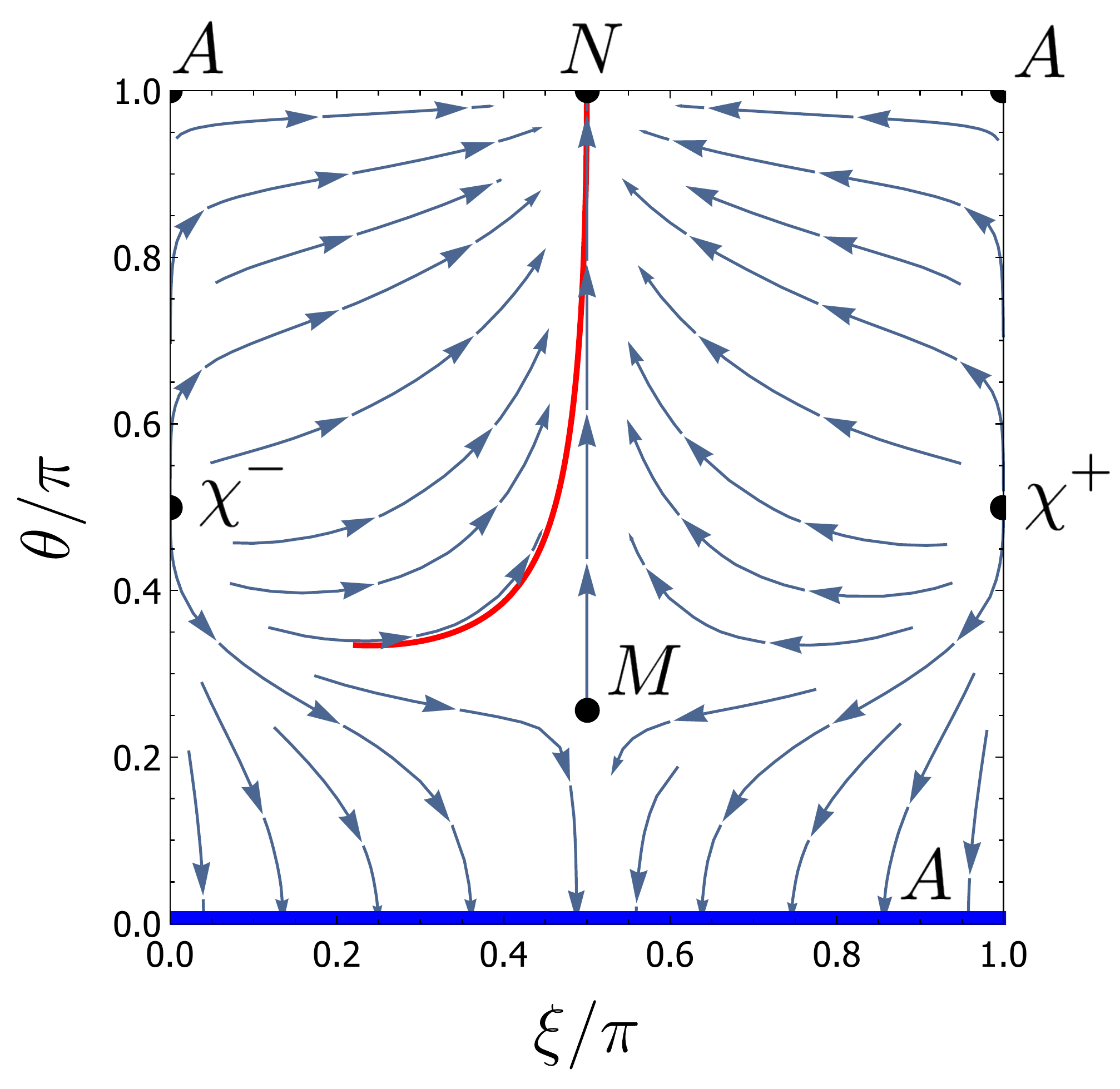} 
\caption{\label{fig:rgflowangle}
RG flows (blue arrows) demonstrate the simultaneous existence of the stable fixed line (blue line) at point A and stable point N, depending on the initial conditions $(\xi, \theta)$ for electron interaction $g=0.72$ ($K=0.4$). Red line corresponds to the RG flow plotted in terms of the conductances   in Fig.~\ref{fig:ConductancesCurve}b.}
\end{figure}

Now we go beyond the weak tunneling approximation and discuss the general RG Eqs.\ \eqref{eq:rgtx} for $K<1$. In this rather complicated case   we have five universal FPs  in the $(\theta,\xi)$-plane, one interaction dependent (i.e. non-universal) fixed point $M$ $(\xi_M=\pi/2, \theta_M=\arccos \frac{1}{3} \left(\sqrt{q^2+6 q-3}-q\right)$ and the lines of fixed points (  $\xi \in (0,\pi)$ and $\theta=0$), see Fig.~\ref{fig:rgflowangle} and Tab.~\ref{tab:universalFP}.  The  points $(\xi=0,\theta=\pi)$, $(\xi=\pi,\theta=\pi)$ and  the fixed line $\theta=0$ correspond to one point $A$ in terms of conductances. 

\begin{table}
\caption{\label{tab:universalFP} The position of the universal fixed points. }
\begin{center}
\begin{tabular}{c|cccccc}
  $\theta$ &$ 0$  & $\pi $& $\pi$ &  $\pi$ & $\pi/2$ &$ \pi/2$  \\ 
  $\xi$ & arb.  &$ 0 $&$ \pi$ & $ \pi/2$ &$ 0$ & $\pi  $\\ \hline
  $G_a$   &$ 1  $&$1 $ & $1$ & $ $0$ $& $3/4$ & $3/4$  \\ 
  $G_b$   & $0$  & $0 $& $0$ & $0$ & $1$ & $1$  \\ 
  $G_{ab}$ & $0 $ &$ 0$ &$ 0 $& $ 0$ & $1/2$ & $-1/2 $ \\ \hline
        FP      & $A$  & $A$ & $A$ &  $N$ & $\chi^-$ & $\chi^+$   
\end{tabular}
\end{center}
\end{table}

For repulsive electron interaction the fixed line $\theta=0$ (the point $A$) is stable and corresponds to effective detachment of the tip from the edge state. It exists simultaneously with the stable point $N$ which corresponds to full breaking of the junction. The line in $(\xi, \theta)$ plane which separates these two basins of attraction connects the unstable points $\chi ^ \pm$ with the saddle point $M$ (see Fig.~\ref{fig:rgflowangle}).  The   chiral FPs  $\chi ^ \pm$ were discussed in detail in \cite{Oshikawa2006,Aristov2013} and they become stable for attractive interaction.  

The whole plane $(\theta,\xi)$ is divided into two regions in Fig.~\ref{fig:rgflowangle}, depending on the interaction strength. One region corresponds to a basin of attraction to the point $N$, the other to the point $A$. The situation is similar to purely tunneling case  (PTC) for symmetric non-chiral $Y$-junction \cite{Aristov2010}. It was shown there that for PTC  the points $A$ or $N$ are stable and separated by the unstable $M$ point.   Any deviation from PTC, which means the backscattering in the main wire, leads to an additional RG flow driving the junction to truly stable $N$ point while $A$ becomes of saddle point type. In our present model we have the chiral PTC junction, and any additional modification of $S$-matrix is forbidden due to topological character of edge state and the absence of backscattering in the main wire. 

Notice that  for vanishing interaction, $g$, the FPs $A$ and $M$ merge, and the basin of attraction for $A$ point vanish. More precisely, $A$ point becomes of saddle-point type in this limit and is stable only for parallel polarization of the tip, $\xi = 0, \pi$.  All RG trajectories in this limit obey  the equation 
\begin{equation}
\cos \xi \tan^2 (\theta/2) = \mbox{const}.
\end{equation}

\begin{figure}
\includegraphics[width=0.85\columnwidth]{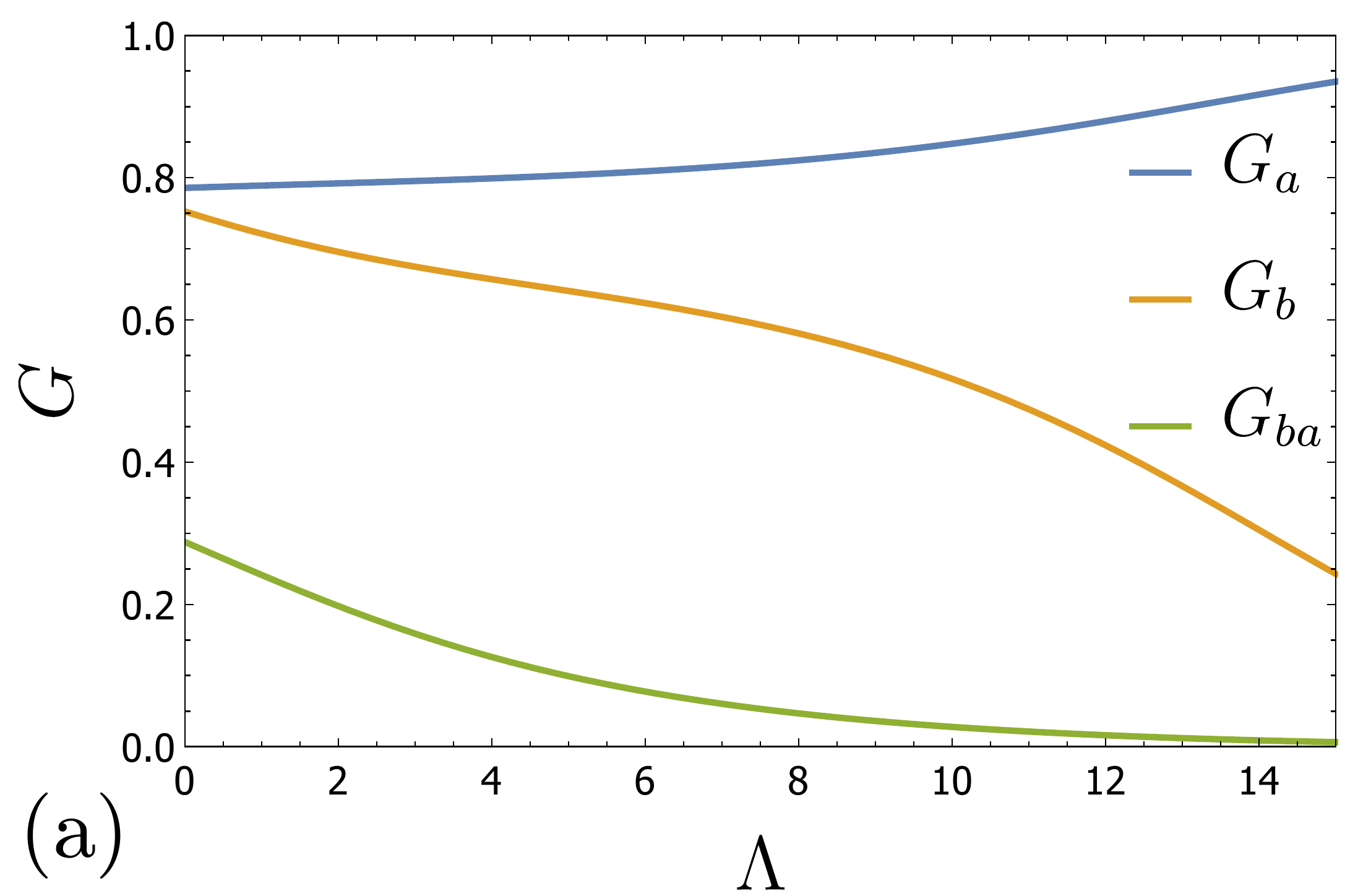} 
\includegraphics[width=0.85\columnwidth]{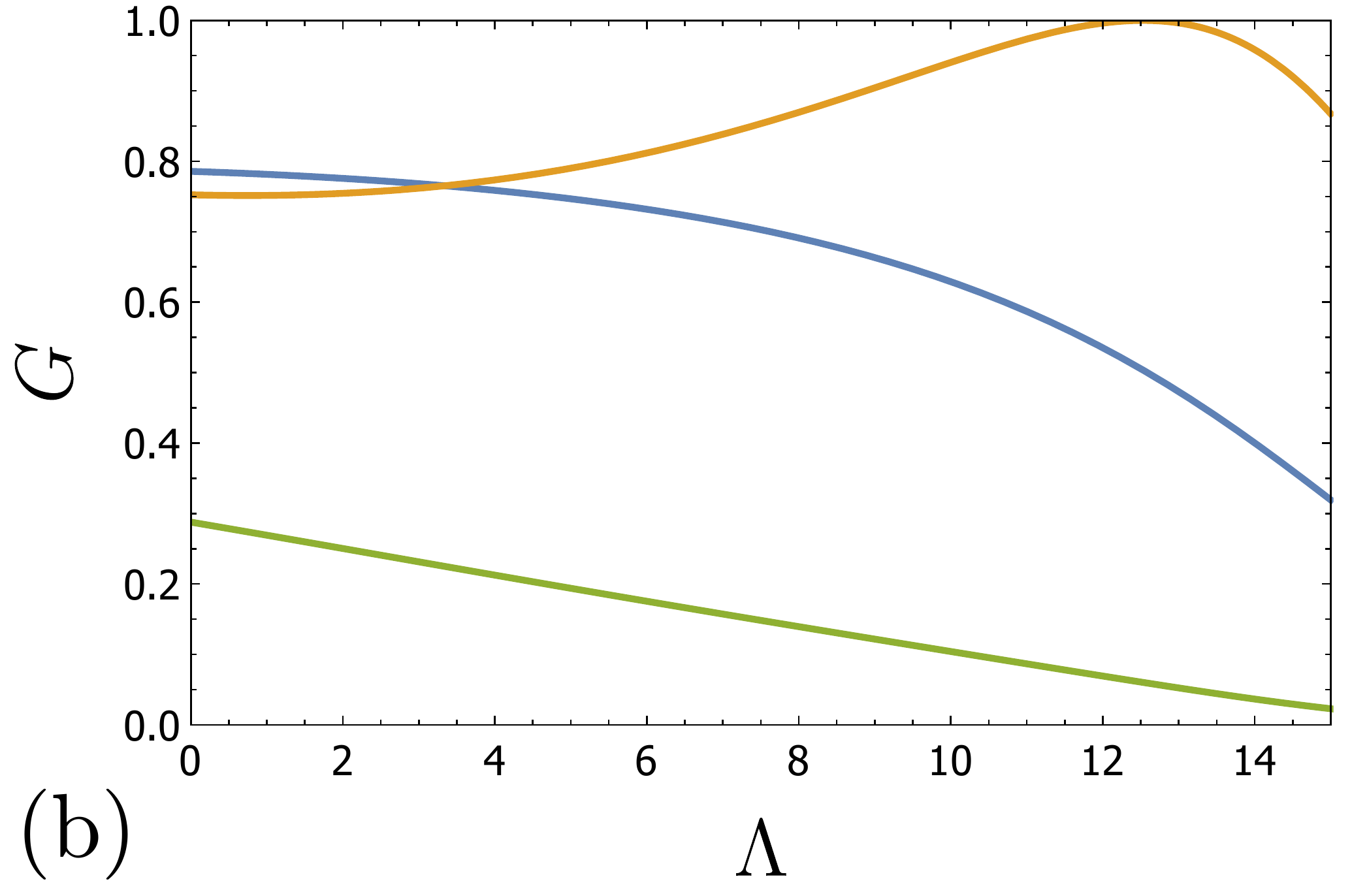} 
\caption{\label{fig:ConductancesCurve}
Full scaling curves for conductances are shown in panel (a) for electron interaction $g=0.47$ ($K=0.6$). The RG flow corresponds to going to the point $A$, i.e. effective detachment of the tip.
The panel (b) shows the RG flows for stronger electron interaction $g=0.72$ ($K=0.4$).  The conductances are renormalized  to the point $N$, i.e. effective full breaking of the junction.  Bare initial conductances on both panels are the same and the qualitative change in the RG flows is caused by the change in the interaction strength.}
\end{figure}
 
Since the position of $M$ point depends on the interaction strength, a qualitatively different behavior of renormalized  conductances may occur even for their equal bare values at different interactions.
 Two examples of full scaling curves for conductances calculated for sizable interaction strength are shown in Fig.~\ref{fig:ConductancesCurve}. The values of bare conductances are the same, but   the panel (a) demonstrates RG flows going to the point $A$, whereas the panel (b) shows the system driven to the point $N$. This difference stems from the fact that changing interaction strength shifts the separating curve between the basins of attraction of the corresponding FPs.

\section{Conclusions \label{sec:conclu}} 

We analyzed the simultaneous renormalization of  asymmetry and total tunneling conductance from fully polarized tip into the helical edge state. In particular we showed that the RG flow for these two quantities  is sensitive to the appearance of non-universal  FP with asymmetry  $\cos\xi_M=0$ and tunneling conductance  
$G_{b} \simeq \theta_M^{2} \simeq 2g$ for small interaction, $g$. 

Our method assumes the existence of  Fermi leads, which are needed for consistent definition of the S-matrix describing the Y-junction in terms of asymptotically free states. Experimentally it can be realized by the  extended leads over the helical edge state.   
The helical edge state, on the other hand, is a topological object which cannot have disconnected ends. The dc current injected from the tunneling tip into the edge state may exit  at another drain Y-junction. This implies the ring geometry with potentially important interference effects.  The above picture of the renormalization of the single Y-junction should be valid when the temperature correlation length $\xi = v_{F}/T$ is smaller than the circumference of the edge state multiplied by $|r_{1}|^{2}$. The importance of interference effects is then reduced  \cite{Dmitriev2010} but the corresponding analysis is beyond the scope of this study.

\acknowledgements

We are grateful to P. W\"olfle, D.B. Gutman and V.Yu. Kachorovskii for   useful discussions. This work was partly supported by RFBR grant No 15-52-06009. The work of D.A. was supported by Russian Science Foundation Grant (Project No. 14-22-00281).

\appendix

\section{ Rotation of polarization axes \label{app:rot}}

Our setup depicted in Fig.\ \ref{fig:setup}   is  characterized by the angle $\xi$ between spin quantization axis in  helical edge states and polarization direction in the tip. This junction is described by the $S$-matrix~\eqref{eq:smatrix}. 
We show that the two-parametric $S$ matrix
 can be obtained from one-parametric one by rotation of the edge states basis.
 
  The Kramers spinor of scattered states is given by  rearrangement of out- states, $\psi_{1,out} \leftrightarrow \psi_{2,out}$, by the matrix 
\begin{equation}
F=\begin{pmatrix}
 0 & 1 & 0 \\
 1 & 0 & 0 \\
 0 & 0 & 1 
\end{pmatrix}.
\end{equation}
The rotation of spin quantization axes is described by 
\begin{equation}
\begin{aligned}
U 
&=
 \begin{pmatrix}
e^{\frac{1}{2} i (\alpha +\gamma )} \cos \frac{\beta }{2} & e^{-\frac{1}{2} i (\alpha -\gamma )} \sin \frac{\beta }{2} &0\\
 -e^{\frac{1}{2} i (\alpha -\gamma )} \sin \frac{\beta }{2} & e^{-\frac{1}{2} i (\alpha +\gamma )} \cos \frac{\beta }{2} & 0 \\
 0 & 0 & 1 
\end{pmatrix} ,
\end{aligned}
\end{equation}
where the upper left 2$\times$2 matrix is $e^{i \gamma \sigma_3 /2}  e^{i \beta \sigma_2 /2}  e^{i \alpha \sigma_3 /2}$ 
with $\alpha$,  $\beta$,  $\gamma$ are Euler angles and $\sigma_{2,3}$ Pauli matrices. For injected spin polarization parallel to the helical edge state, $\xi=0$, the form of $S$ matrix is simplified :
\begin{equation}
S_\uparrow=
 \begin{pmatrix}
 0 & -\cos \theta & \sin \theta  \\
 -1 & 0 & 0 \\
 0 & \sin \theta & \cos \theta 
\end{pmatrix} .
\end{equation}
If   $\xi \neq 0 (\pi)$ then we have to rotate the quantization basis of the helical edge states:
\begin{equation}
S=U S_\uparrow F U^\dagger F.
\label{app-S}
\end{equation}
The dependence of rotated $S$-matrix on angles $\alpha$ and $\gamma$ is removed by additional rephasing. The matrix \eqref{app-S} coincides with ~\eqref{eq:smatrix} after the substitution $\beta \rightarrow - \xi$.


In our paper ~\cite{Aristov2016a} the tunneling from the  unpolarized tip into the helical  edge states was discussed. This X junction was modeled by 4$\times$4 S matrix since there are two spin channels in unpolarized case. It was shown that a certain ratio of the conductances is not renormalized, see equation (26) in \cite{Aristov2016a}.  Applying the above arguments to the situation with X-junction,  we confirm that the discussed ratio corresponds to the angle between the quantization axes for helical edge states  and the tip (the quantity $-2\gamma$ in \cite{Aristov2016a} is the above Euler angle $\beta$). This observation clarifies 
the scale invariance of the ratio, as the choice for the spin quantization basis can be made arbitrarily without changing physical consequences. 

The case of the corner X-junction between two helical edge states~\cite{Aristov2016a} can also be discussed in terms of quantization axes rotation. Three stable FPs were obtained for repulsive interaction. One of them corresponds to the detachment of helical edge states independent of directions of the spin quantization axes. Two other FPs correspond to cases of parallel ($\beta=0$) or antiparallel ($\beta=\pi$) orientation of the quantization axes in individual edges. The choice between these FPs is defined by the value of relative orientation without interaction, $\beta < \pi/2$ or $\beta > \pi/2$, respectively.

\section{Ultraviolet cutoff~\label{app:cutoff}}

The logarithmic corrections to conductance come from consideration of the interaction term, $g \int dx \,dy\, f(x-y)\rho(x) \tilde \rho(y)$, with the range function $f(x-y)$ of the screened Coulomb interaction having the property  $\int dx\, f(x) =1$. In case of Luttinger model the interaction is point-like and  $f(x-y)=\delta(x-y)$, which leads to the Hamiltonian~\eqref{Ham}. However this Luttinger form of interaction leads to ultraviolet divergence which is eventually cut off by another model assumption, formulated in the first line of Eq.~\eqref{Ham} in the main text, namely, the absence of interaction in the vicinity of the junction on the scale $\ell$. We show now that the latter model assumption is not necessary if one assumes the finite range of the interaction, $r_{0}$.

To see it, we recall that the logarithmic type of integral (in the linear response) corresponds to the loop integration of the Green's function $\delta I = \int_{0}^{\infty} d\omega \int_{0}^{L} dx \,dy\, f(x-y)   \sin \omega(y+x)$, see Eq. (19) in Ref.\ \cite{Aristov2014}.  Evaluating this integral we have
 \begin{equation}
\begin{aligned}
\delta I & \simeq \tfrac12 \int_{0}^{\infty} d\omega  \int_{-\infty}^{\infty} dr\, f(r)   \int_{|r|}^{2L} dR\, \sin \omega R    \\
 &  \simeq \int_{0}^{\infty} \frac{d\omega}{2\omega}
 \int_{-\infty}^{\infty} dr\, f(r)   (\cos \omega r - \cos 2\omega L)  \\
\end{aligned}
\end{equation}
It is clear that the finite range of interaction, $r_{0}$, becomes important in the limit of large $\omega \gtrsim r_{0}^{-1}$, cutting off the logarithmic divergence. Therefore $r_{0}$ plays the same role as the model scale $\ell$ and for our purposes we can take the larger of them as the ultraviolet scale of the theory.
 
\section{Fixed point $A$}

\begin{figure*}
\includegraphics[width=0.86\columnwidth]{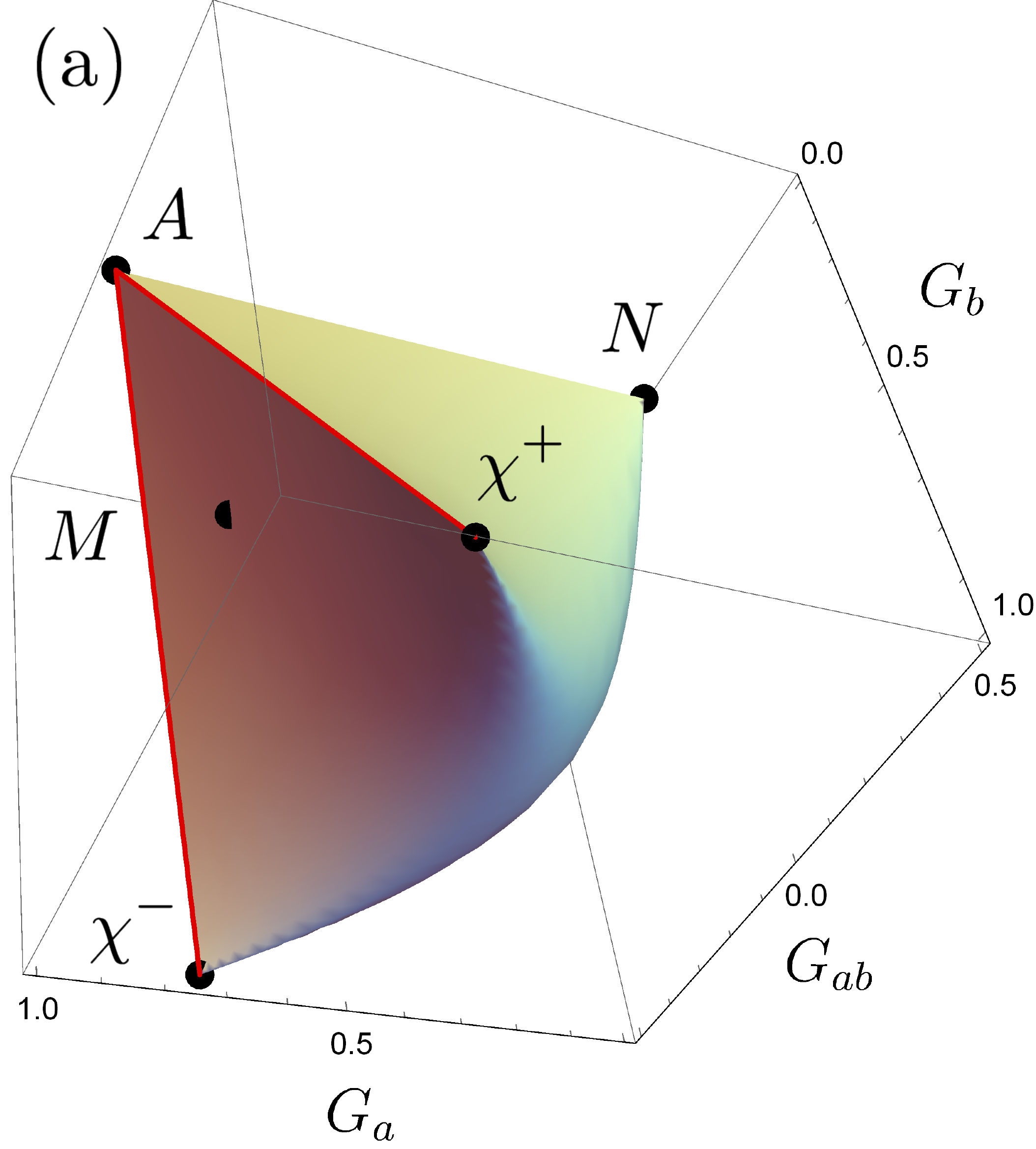} 
\includegraphics[width=0.79\columnwidth]{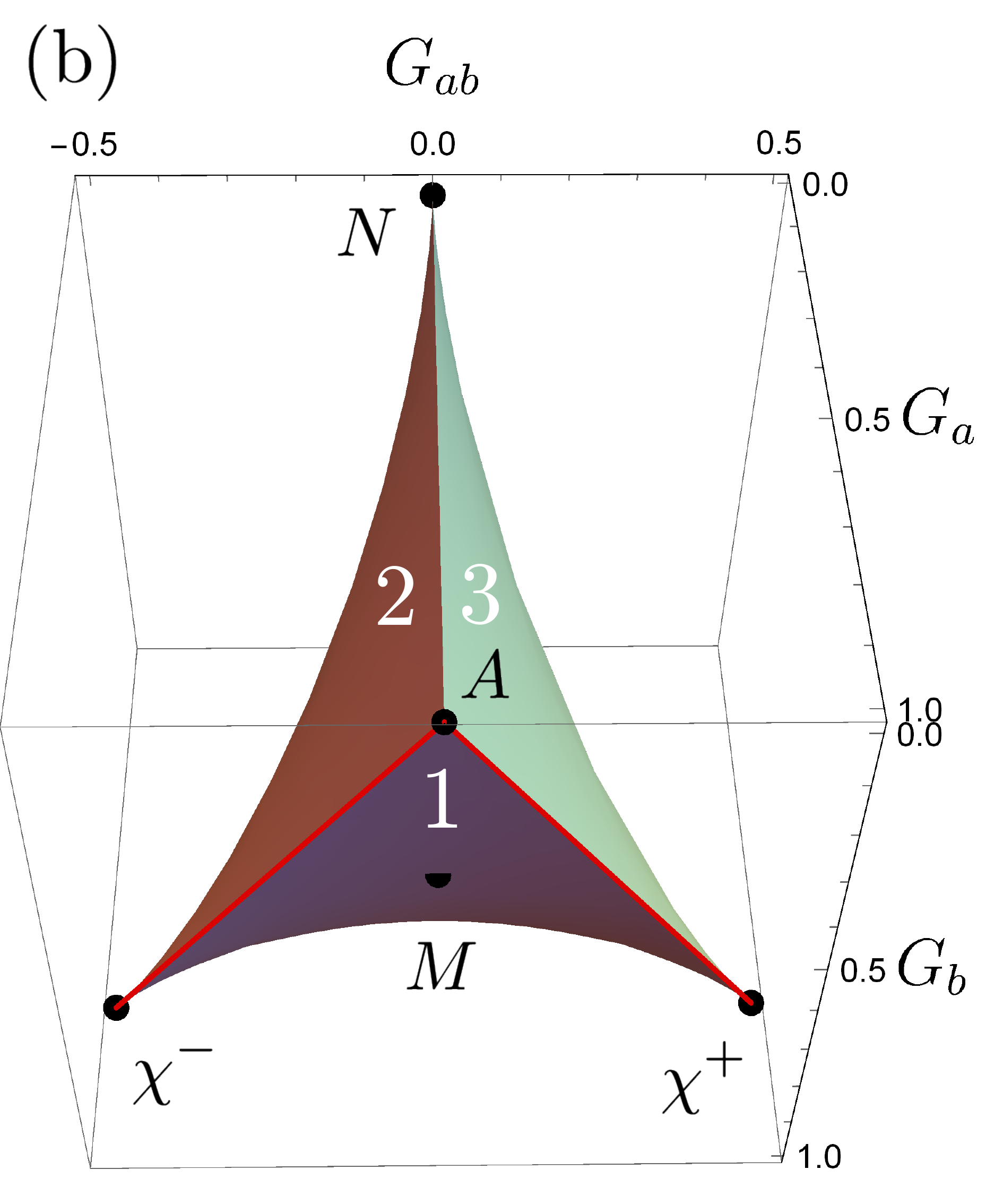} 
\caption{\label{fig:condbody}
2D fixed surface of physical conductances~\eqref{eq:physcond} in 3D conductances space is shown. Fixed points are depicted by black dots. Fixed lines corresponding to $\xi=0$, $\theta \in (0,\pi)$ and $\xi=\pi$,  $\theta \in (0,\pi)$ are shown in red. The panel (b) shows that  three domains   near the point A are separated by ridges, which cannot be crossed by any RG flow.   This explains why  the point $A$ is stable  for repulsive electron interaction in the domain 1  ( $\theta \approx 0$ ), and at the same time unstable in domains 2 and 3    ($\theta \approx \pi$,  $\xi \approx 0 (\pi)$).
}
\end{figure*}

 Renormalization group (RG) analysis of the conductances~\eqref{eq:physcond} was performed in terms of parameters $\theta$ and $\xi$ because the RG equations are more transparent in this form. 
However, one may ask why the same point $A$ described as $\theta=0$ is stable and at the same time unstable when parametrized by $\theta=0$,  $\xi=0(\pi)$. This situation is clarified by considering the surface of conductances in Fig.~\ref{fig:condbody}. It is seen that domains of different character  (stable and unstable) near the point $A$ are separated by three ridges. Two of them are the lines $\xi=0$, $\theta \in (0,\pi)$ and $\xi=\pi$,  $\theta \in (0,\pi)$ depicted by the red lines on the conductances surface,  the third ridge connects $A$ and $N$ points. The domain 1 corresponds to   $\theta \approx 0$ and the domain 2 (3) corresponds to   $\theta \approx \pi$ but $\xi \approx 0 (\pi)$. All these ridges are RG fixed lines, and any RG flow starting on these lines continues along them. It also means that any RG flow cannot pass across these ridges.

For completeness we provide here the scaling exponents at the FPs $A$ and $N$.  The scaling exponent for the point $A$  depends on the direction of approach to this FP as explained above.  In the domain 1  in Fig.\ \ref{fig:condbody} i.e. at $\theta \simeq 0$  we have only one scaling exponent for $\theta$, equal to $(1-K)^{2}/ 4K$. In the domains 2 and 3   in Fig.\ \ref{fig:condbody}, i.e. at  the points $\theta =\pi$, $\xi=0 (\pi)$ we have additional scaling exponent, the RG flows away from $A$ point  along $\xi$ with a weak impurity exponent $(1-K)$. Near the stable $N$ point the exponents along $\theta$ and $\xi$ are $\tfrac12(K^{-1}-1)$ and $(K^{-1}-1)$.


 \end{document}